\pdfoutput=1
\documentclass[conference]{IEEEtran}

\usepackage[utf8]{inputenc}
\usepackage[T1]{fontenc}
\usepackage{cite}
\usepackage{amsmath,amssymb}
\usepackage{graphicx}
\usepackage{textcomp}
\usepackage{xcolor}
\usepackage{booktabs}
\usepackage{multirow}
\usepackage{url}
\usepackage{hyperref}
\hypersetup{
  colorlinks=true,
  linkcolor=black,
  citecolor=blue!60!black,
  urlcolor=blue!60!black
}
\usepackage{balance}
\usepackage{tikz}
\usetikzlibrary{arrows, shapes.geometric, positioning, shadows, patterns, fit, calc, backgrounds}

\newcommand{\AV}{Agile~V}
\newsavebox{\modelcostbox}

\begin{document}

\title{Agile V: A Compliance-Oriented Framework for\\ AI-Augmented Engineering
       --- From Concept to Audit-Ready Delivery}

\author{
  \IEEEauthorblockN{Christopher Koch}
  \IEEEauthorblockA{Agile-V.org\\Germany\\
  \url{https://agile-v.org}}
  \and
  \IEEEauthorblockN{Joshua A.\ Wellbrock}
  \IEEEauthorblockA{Agile-V.org\\Germany\\
  \url{https://agile-v.org}}
}

\maketitle

\begin{abstract}
Current AI-assisted engineering workflows lack a built-in mechanism to maintain
task-level verification and regulatory traceability at machine-speed delivery.
\AV{} addresses this gap by embedding independent verification and audit
artifact generation into each task cycle.  The framework merges Agile iteration
with V-Model verification into a continuous \emph{Infinity Loop}, deploying
specialized AI agents for requirements, design, build, test, and
compliance---governed by mandatory human approval gates.  We evaluate three
hypotheses: (H1)~audit-oriented artifacts emerge as a by-product of development,
(H2)~100\% requirement-level verification is achievable with independent test
generation, and (H3)~verified increments can be delivered with single-digit
human interactions per cycle.  A feasibility case study on a
Hardware-in-the-Loop system ($\approx$500~LOC, 8~requirements, 54~tests)
supports all three hypotheses: audit-oriented documentation was generated
automatically (H1), 100\% requirement-level pass rate was achieved (H2), and
only 6~prompts per cycle were required (H3)---yielding an \textbf{estimated
10--50$\times$ cost reduction} versus a COCOMO~II baseline (sensitivity range
from pessimistic to optimistic assumptions).
These results demonstrate feasibility on a bounded, well-defined project;
generalizability to larger or more ambiguous systems remains to be established
through independent replication.
\end{abstract}

\begin{IEEEkeywords}
Agile, V-Model, AI-augmented engineering, verification, traceability,
quality management, regulatory compliance, open standard, GMP, GxP,
ISO~9001, ISO~13485, GAMP~5, 21~CFR Part~11
\end{IEEEkeywords}

\section{Introduction}\label{sec:intro}

AI coding assistants have moved from pilot programs toward broad organizational adoption~\cite{dellacqua2023jagged,stackoverflow2025ai}.
Engineering teams now routinely generate code, tests, and documentation at a
pace that was unthinkable two years ago~\cite{peng2023copilot}.  Yet this acceleration exposes a
structural weakness: \emph{current AI-assisted engineering workflows are missing
a built-in mechanism to maintain task-level verification and regulatory
traceability at machine-speed delivery}~\cite{he2026cursor,fitzgerald2013scaling,poth2020regulated}.

The resulting business risk is evident.  Organizations in regulated industries
face audit findings when development artifacts lack traceability~\cite{fitzgerald2013scaling,poth2020regulated,ojameruaye2014compliancedebt,dora2024highlights}.  Product
teams can ship faster, but deferring compliance work can accumulate compliance
debt that surfaces during
certification or customer audits~\cite{ojameruaye2014compliancedebt,poth2020regulated,fitzgerald2013scaling}.  Quality managers struggle to enforce review
disciplines when the volume of generated work overwhelms human reviewers~\cite{bacchelli2013codereview}.

The two dominant paradigms each solve only half the problem:
\begin{itemize}
  \item \textbf{Scrum}~\cite{scrum2020} optimizes for adaptability and speed
        but provides no built-in mechanism for verification, traceability, or
        regulatory documentation~\cite{fitzgerald2013scaling}---a gap that
        widens further at the pace AI agents produce.
  \item \textbf{V-Model}~\cite{vmodel2005} delivers the rigor and traceability
        regulators expect but is project-bound, phase-gated, and too slow for
        continuous AI-driven delivery.
\end{itemize}

\AV{}\footnote{Agile~V is a trademark of Agile-V.org.} addresses this gap by
embedding independent verification and audit artifact generation into each task
cycle.  It is a unified open standard (CC~BY-SA~4.0~\cite{agilev2026}) that
merges Agile iteration with V-Model verification into a single, repeatable
workflow---the \emph{Infinity Loop}---executed continuously at the task level.
The framework is operationalized through a library of composable AI agent
skills~\cite{agilev_skills2026} that any organization can deploy within
existing toolchains.

This paper presents both a conceptual framework and a feasibility case study.
We evaluate \AV{} against three hypotheses:

\begin{itemize}
  \item \textbf{H1 (Audit-Evidence By-Product):} Using \AV{}'s Infinity Loop with
        human approval gates, a small AI-agent team can produce a structured
        audit-evidence artifact set (requirements specification, traceability
        matrix, test evidence, decision log) as a by-product of development for
        a bounded project.
  \item \textbf{H2 (Verification Pass Rate):} For a bounded system, \AV{} can
        achieve 100\% requirement-level verification pass rate with independent
        test generation.
  \item \textbf{H3 (Minimal Human Interaction):} For a bounded system, \AV{}
        can deliver a working, verified increment with a small number of human
        interactions per cycle (e.g., single-digit prompts) while maintaining
        full traceability.
\end{itemize}

We evaluate these hypotheses through a Hardware-in-the-Loop (HIL) case study
spanning two Infinity Loop cycles.  The study quantifies the business impact:
an estimated 10--50$\times$ cost reduction (versus a COCOMO~II baseline,
depending on assumptions; see Section~\ref{sec:sensitivity}) with structured
audit-evidence documentation produced as a by-product of the engineering workflow.
The second cycle validates the framework's iterative correction mechanism,
showing that AI-generated artifacts can be systematically hardened through
structured change requests.

\section{Related Work}\label{sec:related}

\subsection{Agile and V-Model Hybrids}
Several attempts have been made to reconcile Agile iteration with V-Model
verification.  The W-Model~\cite{spillner2002wmodel} extends the V-Model by
introducing early test activities in parallel with each development phase,
reducing the delay between design and verification.  Yet, the W-Model
remains project-scoped and does not address continuous, task-level verification.
The Scaled Agile Framework (SAFe)~\cite{safe2024} adds governance layers to
Scrum, including compliance and regulatory tracks, but does not prescribe how
AI-generated artifacts should be verified or how traceability should be
maintained at machine speed.  \AV{} differs from both by operating at the
\emph{individual task} level and by embedding verification as a structural
constraint of the workflow rather than as an optional overlay.

\subsection{DevSecOps and Automated Compliance}
DevSecOps pipelines~\cite{myrbakken2017devsecops} integrate security and
compliance checks into CI/CD workflows, typically through static analysis,
container scanning, and policy-as-code gates.  While these pipelines automate
\emph{execution} of predefined checks, they do not generate the compliance
artifacts themselves (e.g., requirements specifications, traceability matrices,
decision rationale logs).  \AV{} complements DevSecOps by producing these
artifacts as a workflow by-product, enabling organizations to adopt both
approaches: DevSecOps for infrastructure-level controls and \AV{} for
engineering-level verification and documentation.

\subsection{AI-Assisted Software Engineering}
Recent empirical studies have measured the productivity impact of AI coding
assistants.  Peng et al.~\cite{peng2023copilot} reported a 55.8\% task
completion speedup with GitHub Copilot, while
Dell'Acqua et al.~\cite{dellacqua2023jagged} found that AI-assisted
consultants produced 40\% higher quality work---but only within the model's
capability boundary.  He et al.~\cite{he2026cursor} demonstrated that
unstructured AI-assisted development increases code complexity and technical
debt over time.  These findings collectively motivate \AV{}'s design: the
framework captures the productivity gains documented by Peng et al.\
and Dell'Acqua et al.\ while mitigating the quality risks identified by
He et al.\ through mandatory verification gates and independent test
generation.

\subsection{LLM-Based Test Generation}
Recent work uses LLMs to generate tests from \emph{code}: CodaMosa~\cite{lemieux2023codamosa}
combines search-based generation with LLM-produced seeds to improve branch
coverage, and ChatUniTest~\cite{chen2024chatunitest} achieves competitive
coverage when guided by focal-method context.  These tools maximize
\emph{code-level} coverage.  \AV{}'s Test Designer differs by generating tests
from \emph{requirements only}, ensuring structural independence between build
and verification.  The approaches are complementary: LLM-based coverage tools
could augment \AV{}'s requirement-level suite in future work.

\subsection{Spec-Driven AI Development Tools}
A growing class of open-source tools addresses context management and task
orchestration for AI coding agents.  Representative examples include Get Shit
Done (GSD)~\cite{gsd2026}, BMAD, and SpecKit---meta-prompting systems that
decompose projects into phases, spawn parallel sub-agents, and manage context
window limits.  These tools are \emph{productivity tooling}, not \emph{process
frameworks}: they optimize how an AI agent builds code but prescribe neither
independent verification (build and test share context), nor regulatory
traceability (no requirement-to-test mapping or audit trail), nor human
governance gates.

\AV{} operates at a different abstraction level: where GSD is a \emph{build
accelerator}, \AV{} is an \emph{engineering process} that ensures artifacts are
verifiable, traceable, and structured for audit review.  In v1.3~\cite{agilev_skills2026}, we
integrated GSD's execution-layer patterns into the skills library with one
constraint: all GSD-derived mechanisms operate \emph{within} the Infinity
Loop's verification boundaries, demonstrating that productivity tooling and
process governance can be composed.

\section{The \AV{} Framework}\label{sec:framework}

\subsection{Core Principles}\label{sec:manifesto}

\AV{} is built on four operational principles that redefine how engineering
organizations use AI while maintaining the governance that regulators, auditors,
and customers demand:

\begin{enumerate}
  \item \textbf{Verified Iteration} over Unchecked Velocity ---
        Every AI-generated artifact is tested before it advances.  Speed is a
        by-product of confidence, not a substitute for it.
  \item \textbf{Traceable Agency} over Opaque Autonomy ---
        The audit trail records \emph{who}---human or AI agent---made each
        decision and \emph{why}, satisfying the accountability requirements of
        ISO~9001~\cite{iso9001} and GxP regulations.
  \item \textbf{Living Compliance} over Static Documentation ---
        Regulatory documentation is generated as a by-product of the
        engineering workflow, eliminating the post-hoc ``documentation sprint''
        that delays releases.
  \item \textbf{Human Curation} over Human Execution ---
        Engineers shift from writing code to directing intent, reviewing
        designs, and approving releases.  AI handles the execution; humans
        retain authority.
\end{enumerate}

\subsection{The Infinity Loop Workflow}\label{sec:loop}

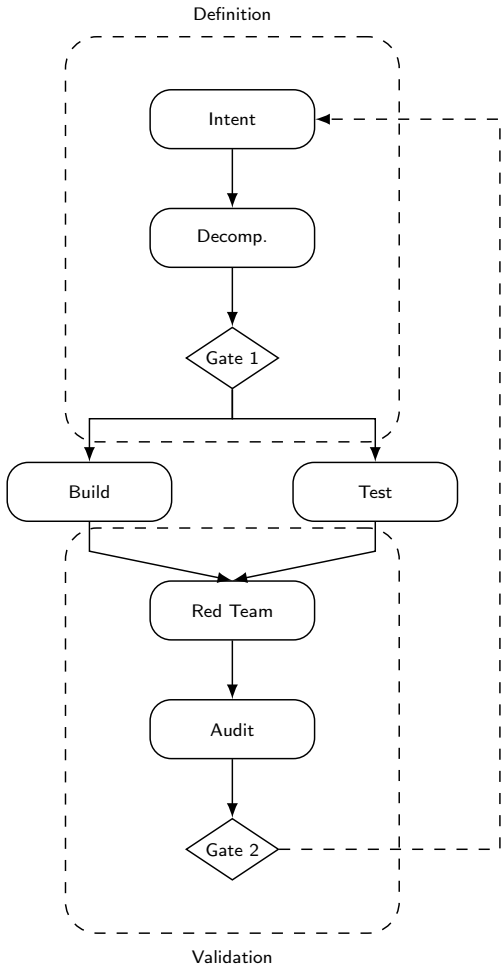
\begin{figure}[t]
\centering
\resizebox{0.8\columnwidth}{!}{
\begin{tikzpicture}[
    scale=0.7,
    transform shape,
    >=latex,
    node distance=1.5cm,
    auto,
    font=\footnotesize\sffamily,
    state/.style={
        rectangle, 
        rounded corners=2mm, 
        draw=black, 
        fill=white, 
        align=center, 
        minimum height=0.8cm, 
        text width=2.0cm,
        font=\scriptsize\sffamily
    },
    gate/.style={
        diamond, 
        aspect=1.5,
        draw=black, 
        fill=white, 
        align=center, 
        inner sep=0pt,
        text width=1.0cm,
        font=\scriptsize\sffamily
    },
    line/.style={
        draw, 
        ->, 
        color=black
    }
]

    % Coordinates for the Infinity Loop
    
    % Left Loop Nodes (Intent & Decomposition)
    \node [state] (intent) {Intent};
    \node [state, below=0.8cm of intent] (decomp) {Decomp.};
    \node [gate, below=0.8cm of decomp] (gate1) {Gate 1};

    % Center (Synthesis) - Split to sides to form loop
    \node [state, below left=1.2cm and 0.5cm of gate1] (build) {Build};
    \node [state, below right=1.2cm and 0.5cm of gate1] (test) {Test};

    % Right Loop Nodes (Verification) - Converging back
    \node [state, below=2.6cm of gate1] (verify) {Red Team};
    \node [state, below=0.8cm of verify] (compliance) {Audit};
    \node [gate, below=0.8cm of compliance] (gate2) {Gate 2};

    % Phase grouping boxes (dashed outline, matching hil_arch style)
    \node [fit=(intent)(decomp)(gate1), 
           draw=black,
           dashed,
           rounded corners=3mm,
           inner xsep=0.8cm,
           inner ysep=0.5cm] (defgroup) {};
    \node [above=0.1cm of defgroup, font=\scriptsize\sffamily] {Definition};
    
    \node [fit=(verify)(compliance)(gate2), 
           draw=black,
           dashed,
           rounded corners=3mm,
           inner xsep=0.8cm,
           inner ysep=0.5cm] (valgroup) {};
    \node [below=0.1cm of valgroup, font=\scriptsize\sffamily] {Validation};

    % Draw the flow (Vertical Infinity Loop / Hourglass)
    
    % Downward flow
    \draw [line] (intent) -- (decomp);
    \draw [line] (decomp) -- (gate1);
    
    % Split
    \draw [line] (gate1.south) -- +(0,-0.4) -| (build.north);
    \draw [line] (gate1.south) -- +(0,-0.4) -| (test.north);
    
    % Merge
    \draw [line] (build.south) |- +(0,-0.4) -- (verify.north);
    \draw [line] (test.south) |- +(0,-0.4) -- (verify.north);
    
    % Continue down
    \draw [line] (verify) -- (compliance);
    \draw [line] (compliance) -- (gate2);
    
    % Feedback / Iteration (Closing the loop)
    % Large loop back to top - Wide arc to clear all nodes
    \draw [line, dashed] (gate2.east) -- +(3.0,0) |- (intent.east);

\end{tikzpicture}
}
\caption{The Agile V Infinity Loop: Definition (top) flows through Synthesis (center) to Validation (bottom), with a feedback path closing the loop.}
\label{fig:infinity_loop}
\end{figure}

The central mechanism of \AV{} is the \emph{Infinity Loop}
(Fig.~\ref{fig:infinity_loop}), a continuous workflow that replaces sequential
project phases with a repeatable cycle executed on every task:

\subsubsection{Definition --- Intent \& Decomposition}
An engineer provides high-level \emph{Product Intent} (e.g., ``build a HIL test
system for this analyzer'').  The \textbf{Requirement Architect} agent
decomposes it into atomic, traceable requirements with explicit acceptance
criteria.  The \textbf{Logic Gatekeeper} validates feasibility against physical
and organizational constraints.  \textbf{Human Gate~1} ensures the team agrees
on scope before any code is written.

\subsubsection{Apex --- Synthesis}
Two agents work \emph{in parallel} to prevent confirmation bias---a key
differentiator from traditional code-then-test workflows:
\begin{itemize}
  \item The \textbf{Build Agent} generates deliverables (code, schematics,
        configuration).
  \item The \textbf{Test Designer} creates a verification suite from the
        \emph{requirements only}---never from the code---ensuring independent
        test coverage.
\end{itemize}

\subsubsection{Validation --- Verification \& Compliance}
The \textbf{Red Team Verifier} executes the test suite against the build
artifacts and produces an objective pass/fail report.  The \textbf{Compliance
Auditor} captures the full decision rationale---\emph{why} each design choice
was made---generating structured audit-evidence logs in real time.  \textbf{Human Gate~2}
gives the responsible engineer final approval authority before release.

\subsection{Context Engineering}\label{sec:context}

A practical challenge for any AI-agent-based workflow is \emph{context window
degradation}: as a session accumulates tokens, model reasoning quality
declines~\cite{gsd2026}.  \AV{} v1.3 addresses this by aligning context
boundaries with V-Model positions.  Definition-phase agents read requirement
files directly; Synthesis agents (Build Agent, Test Designer) receive
requirement IDs and paths but operate in separate, fresh contexts;
Validation-phase agents (Red Team Verifier) never inherit Build Agent context,
preserving the independence required by the Red Team Protocol.  A thin orchestrator ($\approx$10--15\% of
the context window) coordinates agents, passing file references rather than file
contents, and spawning fresh sub-agent contexts so that no single session
exceeds 50\% of the available window.  Independent tasks are parallelized in
\emph{waves} with fresh context per agent---enabling the Build Agent and Test
Designer to synthesize concurrently without shared state.

\subsection{Persistent Memory}\label{sec:persistent_memory}
Agent sessions start without implicit repository knowledge, so exploration-only
workflows tend to re-discover constraints and miss non-obvious conventions.
\AV{} therefore maintains \emph{persistent memory} as a curated,
version-controlled store of high-signal project knowledge (entrypoints,
commands, invariants, and decision rationales) that tells agents \emph{where to
look} and what must not change.

To keep active context small, long-lived knowledge is retrieved on demand rather
than re-injected wholesale~\cite{lewis2020rag}, using bounded working-memory
strategies to fit within a session~\cite{packer2023memgpt}.  Persistent memory
should store non-sensitive summaries and file pointers rather than secrets or
proprietary raw logs.

\section{Case Study: HIL Test System}\label{sec:casestudy}

To demonstrate \AV{} on a representative industrial project, the framework was
applied to deliver a Python-based Hardware-in-the-Loop (HIL) test environment
for a Saleae Logic Analyzer---the type of system commonly required in
electronics manufacturing, automotive validation, and medical device
verification.

\subsection{Project Scope}
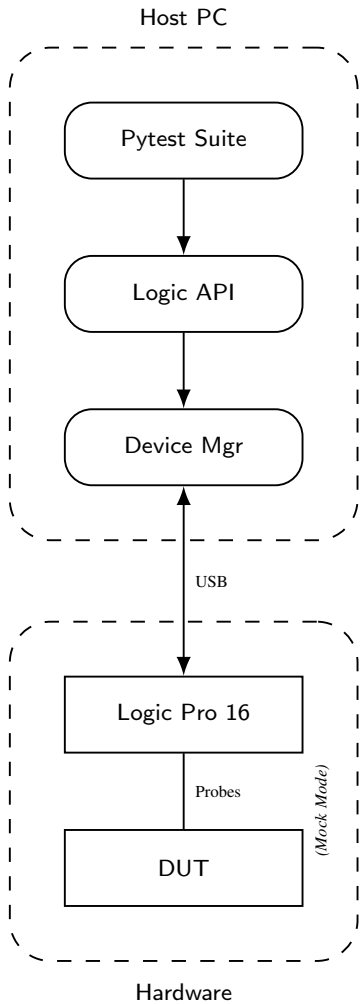
\begin{figure}[!t]
\centering
\resizebox{0.6\columnwidth}{!}{
\begin{tikzpicture}[
    scale=0.7,
    transform shape,
    >=latex,
    node distance=1.5cm,
    auto,
    font=\small\sffamily,
    component/.style={
        rectangle, 
        draw=black, 
        fill=white, 
        align=center, 
        minimum height=0.8cm, 
        minimum width=2.5cm,
        rounded corners=2mm,
        font=\scriptsize\sffamily
    },
    hardware/.style={
        rectangle, 
        draw=black, 
        fill=white, 
        align=center, 
        minimum height=0.8cm, 
        minimum width=2.5cm,
        font=\scriptsize\sffamily
    },
    group/.style={
        rectangle, 
        draw=black, 
        dashed, 
        inner sep=0.4cm, 
        rounded corners=3mm
    },
    line/.style={
        draw, 
        ->, 
        color=black
    }
]

    % Nodes - Vertical Stack for Single Column
    
    % Host PC Software
    \node [component] (script) {Pytest Suite};
    \node [component, below=0.8cm of script] (wrapper) {Logic API};
    \node [component, below=0.8cm of wrapper] (manager) {Device Mgr};
    
    % Hardware - Below Host PC
    \node [hardware, below=2.0cm of manager] (saleae) {Logic Pro 16};
    \node [hardware, below=0.8cm of saleae] (dut) {DUT};
    
    % Grouping - Host PC
    \node [group, fit=(script) (wrapper) (manager)] (host) {};
    % Label moved slightly higher to avoid group boundary overlap
    \node [above=0.1cm of host, font=\scriptsize\sffamily] {Host PC};

    % Grouping - Hardware
    \node [group, fit=(saleae) (dut)] (hw) {};
    % Label moved slightly lower to avoid group boundary overlap
    \node [below=0.1cm of hw, font=\scriptsize\sffamily] {Hardware};

    % Connections
    \draw [line] (script) -- (wrapper);
    \draw [line] (wrapper) -- (manager);
    
    % USB Connection (Host to HW)
    \draw [line, <->] (manager.south) -- node [midway, right, font=\tiny] {USB} (saleae.north);
    
    % Probing
    \draw [line, -] (saleae.south) -- node [midway, right, font=\tiny] {Probes} (dut.north);

    % Annotations
    \node [right=0.2cm of dut, font=\tiny\itshape, rotate=90] {(Mock Mode)};

\end{tikzpicture}
}
\caption{System Architecture: Agile V Agents generated the Python Host components (top), integrating with the physical Logic Analyzer (bottom).}
\label{fig:hil_arch}
\end{figure}

The deliverable (Fig.~\ref{fig:hil_arch}) had to meet four operational
requirements typical of production test systems:
\begin{itemize}
    \item Abstraction of device communication (DUT) for multi-product reuse.
    \item Automation of the Saleae Logic~2 software via its
          API~\cite{saleae2026}.
    \item Synchronization of test scripts with hardware capture ($<100$\,ms
          latency).
    \item Jupyter Notebook integration for engineering analysis and reporting.
\end{itemize}

\subsection{Study Design}\label{sec:study_design}
The case study follows a single-case embedded design~\cite{fitzgerald2013scaling}
with two units of analysis (Cycle~1: initial delivery; Cycle~2: change request).
The study was conducted by the framework's authors, which introduces potential
bias (Section~\ref{sec:threats}).

\textbf{Evaluation criteria.}  Each hypothesis maps to observable metrics:
H1 is evaluated by the presence and completeness of six artifact types
(requirements spec, traceability matrix, test log, decision rationale, risk
register, validation summary); H2 by the requirement-level pass rate and the
number of Red Team findings; H3 by the count of human prompts per cycle.

\textbf{Data collection.}  All agent interactions were conducted through
commercial AI platforms (Gemini~1.5~Pro for Cycle~1, Claude~Opus~4.6 for
Cycle~2).  Prompt counts, artifact inventories, and test results were recorded
from the session logs and the \texttt{.agile-v/} state directory.  The COCOMO~II
baseline was estimated independently of the case study execution.

\subsection{Delivery Execution}
The project completed two full passes through the Infinity Loop, demonstrating
both initial delivery (Cycle~1) and iterative hardening (Cycle~2).

\subsubsection{Cycle~1 --- Initial Delivery}
\begin{enumerate}
    \item \textbf{Intent:} The project lead described the HIL system and its
    hardware constraints in plain language.
    \item \textbf{Decomposition:} The Requirement Architect produced 7~formal
    requirements (REQ-0001 to REQ-0007) with measurable acceptance criteria,
    including ``Mock Device'' mode for CI/CD environments without physical
    hardware.
    \item \textbf{Gate~1:} The Logic Gatekeeper verified that Python~3.10+ and
    the \texttt{saleae-automation} SDK were available and compatible.  The
    project lead approved the Blueprint.
    \item \textbf{Synthesis:} The Build Agent delivered
    \texttt{src/hil/} (LogicAnalyzer wrapper, DeviceManager) while the Test
    Designer independently created \texttt{tests/} and \texttt{notebooks/}.
    \item \textbf{Verification:} The Red Team Verifier executed the initial test
    suite, achieving a 100\% pass rate at the requirement level.  However, the
    Cycle~1 suite was narrower than the final 54-test suite: it covered
    the functional contract of each module but lacked the input validation,
    cycle-field, and state-directory tests that Cycle~2 later added.  The
    first-pass success should therefore be read with the caveat that the
    test surface was expanded significantly in the subsequent cycle.
\end{enumerate}

\subsubsection{Cycle~2 --- Change Request and Compliance Hardening}
Cycle~2 was triggered by CR-0001 (Agile~V Skill Upgrade v1.3), which required
structural changes across the codebase:
\begin{enumerate}
    \item \textbf{Intent:} Upgrade to vendor-agnostic device abstraction,
    add multi-cycle traceability, and harden compliance artifacts.
    \item \textbf{Decomposition:} The Requirement Architect added REQ-0008
    (\texttt{.agile-v/} state directory for persistent process state) and
    modified 4~existing requirements to incorporate cycle-aware fields.  The
    total grew from 7 to 8~verified requirements.
    \item \textbf{Gate~1:} The Logic Gatekeeper validated the change scope.
    \item \textbf{Synthesis:} The Build Agent refactored
    \texttt{DeviceInterface} from a vendor-specific SDK binding to a
    transport-agnostic Abstract Base Class, added a \texttt{cycle} field to
    \texttt{TestResult} and the logging subsystem, and introduced a
    GitHub Actions CI pipeline testing Python~3.10--3.13.  The Test Designer
    expanded the suite from the initial set to \textbf{54~automated tests}.
    \item \textbf{Verification:} The Red Team Verifier identified
    \textbf{10~findings} (6~MAJOR, 4~MINOR)---including stale API references,
    a missing \texttt{cycle} field, an async property anti-pattern, and input
    validation gaps.  All MAJOR findings were resolved by the Build Agent;
    the test suite achieved 54/54 pass post-rework.
\end{enumerate}

This second cycle is significant because it demonstrates the Infinity Loop
operating as designed: a change request triggered a controlled iteration, Red
Team findings drove concrete fixes, and full traceability was maintained across
cycles---exactly the corrective-action loop that ISO~9001:2015
Clause~10.2~\cite{iso9001} requires.

\subsection{Operational Metrics}\label{sec:stats}
The complete system---source code, tests, documentation, and
configuration---was delivered across two cycles with minimal engineer
involvement:

\begin{table}[h]
    \centering
    \caption{Session Metrics (After Cycle~2)}
    \label{tab:metrics}
    \begin{tabular}{llr}
        \toprule
        \textbf{Metric} & \textbf{Scope} & \textbf{Value} \\
        \midrule
        Verified Requirements & Cumulative & 8 \\
        Automated Tests & Cumulative & 54 \\
        Source Code & Cumulative & $\approx$500 LOC \\
        User Prompts & Per cycle & 6 \\
        Red Team Findings & C2 only & 10 (all resolved) \\
        CI Matrix & Cumulative & 4 Python versions \\
        \bottomrule
    \end{tabular}
\end{table}

The metrics in Table~\ref{tab:metrics} serve as the primary data source for
evaluating H1--H3 in Section~\ref{sec:results}.

\subsection{Tooling and Infrastructure}\label{sec:compute}

\AV{} requires only commercially available AI platforms.  The case study used
Gemini 1.5~Pro (Cycle~1) and Claude Opus~4.6 (Cycle~2), demonstrating
model-agnosticism empirically---the text-based agent skills are independent of
the underlying model, and \texttt{.agile-v/config.json} tracks which provider
was used per cycle for auditability.  The context engineering mechanisms
(Section~\ref{sec:context}) decompose work into sub-agent scopes of
$\leq$50\% window utilization, making the framework viable on models with
smaller context windows than those used in the case study.

Cycle~2 introduced a \texttt{.agile-v/} state directory (REQ-0008) that
persists process artifacts (configuration, change log, approvals, risk
register, traceability matrix, Red Team report, validation summary) alongside
source code in version control, serving as both agent context for cross-session
continuity and as an evidence bundle for audit review.

\section{Business Case}\label{sec:cost}

To quantify the return on investment, the recorded \AV{} session was compared
against industry-standard benchmarks for a traditional team delivering the same
scope: a HIL test system with ISO~9001:2015~\cite{iso9001} documentation.

\subsection{Baseline: Traditional Delivery}
The traditional estimate is derived from COCOMO~II parametric cost
modeling~\cite{boehm2000cocomo}, calibrated for a quality-critical embedded
test system of moderate complexity:
\begin{itemize}
    \item \textbf{Requirements \& Architecture:} 16~hours (2~days).
    \item \textbf{Implementation:} 40~hours (1~week).
    \item \textbf{Test Engineering:} 24~hours (3~days).
    \item \textbf{Compliance Documentation:} 24~hours (3~days).
    \item \textbf{Total Effort:} $\approx$\,104~hours (2.5~weeks).
    \item \textbf{Estimated Cost:} \$15,600 (at \$150/hr blended rate).
\end{itemize}

This estimate is conservative.  The Standish Group's CHAOS
data~\cite{standish2015} reports that only 29\% of IT projects are delivered on
time and on budget; the remainder experience cost overruns averaging 45\% or
fail outright---suggesting the true expected cost is likely higher.

\subsection{\AV{} Delivery Cost Across Models}
The \AV{} framework is model-agnostic.  To illustrate the cost range,
Table~\ref{tab:model_cost} compares the compute cost of the HIL case study
(estimated at $\approx$500k input tokens, $\approx$25k output tokens based on
session length and average file sizes) across representative commercially
available models.  Gemini~1.5~Pro values report the observed Cycle~1 billed
compute total; all other rows use February 2026 API list pricing
estimates~\cite{anthropic_pricing,google_gemini_pricing,openai_gpt52_pricing}.

\begin{table}[h]
    \centering
    \caption{AI Compute Cost by Model (HIL Case Study)}
    \label{tab:model_cost}
    \sbox{\modelcostbox}{%
    \begin{tabular}{lrrr}
        \toprule
        \textbf{Model} & \textbf{Input} & \textbf{Output} & \textbf{Total} \\
        \midrule
        Gemini 1.5 Pro\textsuperscript{*} & \$1.75 & \$2.63 & \$4.38 \\
        Gemini 2.5 Pro & \$0.63 & \$0.25 & \$0.88 \\
        Claude Sonnet 4.6 & \$1.50 & \$0.38 & \$1.88 \\
        Claude Opus 4.6 & \$2.50 & \$0.63 & \$3.13 \\
        GPT-5 mini & \$0.13 & \$0.05 & \$0.18 \\
        GPT-5.2 & \$0.88 & \$0.35 & \$1.23 \\
        \bottomrule
    \end{tabular}
    }%
    \usebox{\modelcostbox}
    \par\smallskip
    {\footnotesize\parbox{\wd\modelcostbox}{\raggedright\textsuperscript{*}Gemini~1.5~Pro values are the Cycle~1 billed compute total. All other rows are list-price estimates using the same token counts; Gemini~2.5~Pro assumes the standard (<=200k prompt) pricing tier~\cite{google_gemini_pricing}. Pricing varies by provider features (e.g., caching, batching, long-context tiers).\par}}
\end{table}

\smallskip\noindent
Including 4~hours of engineer time at \$150/hr, the \emph{per-cycle} project
cost ranges from \textbf{\$601} to \textbf{\$605} across the models in
Table~\ref{tab:model_cost}.
The AI compute cost is negligible relative to human effort in all
cases---representing less than 1\% of the total.

\textbf{Methodological notes:}
\begin{enumerate}
    \item The \AV{} cost figure captures the direct execution cost (human
    gate-review time plus AI compute) for a \emph{single cycle}.  The case
    study completed two cycles; the total observed cost is approximately
    \$1{,}200 (2~$\times$~4~hours + $\approx$\$10 compute).  The baseline
    25$\times$ figure is based on a single delivery cycle; the sensitivity
    range is 10--50$\times$ (Section~\ref{sec:sensitivity}).
    \item The traditional baseline (\$15{,}600) is a COCOMO~II \emph{estimate},
    not a measured cost.  The comparison is therefore between a parametric
    model and an observed session, which limits the precision of the ratio.
    \item One-time adoption costs (framework learning, skill configuration)
    are excluded and estimated at 8--16~hours for a team's first \AV{}
    deployment.  These costs are amortized across projects.
\end{enumerate}

\subsection{Impact Summary}
\begin{itemize}
    \item \textbf{Cost:} Estimated 10--50$\times$ reduction (\$15,600 COCOMO~II
    baseline $\rightarrow$ \$601--\$605 observed per cycle; 96\% at baseline).
    \item \textbf{Time-to-delivery:} Estimated 25$\times$ acceleration
    (2.5~weeks baseline $\rightarrow$ 4~hours observed per cycle).
    \item \textbf{Compliance coverage:} Higher artifact completeness than
    typical for a 2-week sprint---requirements specification, traceability
    matrix, test logs, and decision rationale all delivered automatically.
    \item \textbf{Audit evidence:} Documentation produced \emph{during}
    development, not weeks later.  No separate compliance retrofit required.
    Whether these artifacts satisfy a specific auditor's requirements depends on
    organizational controls, scope, and regulatory context.
\end{itemize}

These gains are directionally consistent with the productivity improvements
reported by Peng et al.~\cite{peng2023copilot} and
McKinsey~\cite{mckinsey2023genai}, while the Red Team Protocol addresses the
quality risks identified by He et al.~\cite{he2026cursor}
(Section~\ref{sec:related}).

\subsection{Sensitivity Analysis}\label{sec:sensitivity}

The 25$\times$ cost reduction is based on specific assumptions.  To assess
robustness, Table~\ref{tab:sensitivity} varies the four key parameters across
pessimistic, baseline, and optimistic scenarios.

\begin{table}[h]
    \centering
    \caption{Sensitivity Analysis: Cost Reduction Factor}
    \label{tab:sensitivity}
    \begin{tabular}{p{2.1cm}rrr}
        \toprule
        \textbf{Parameter} & \textbf{Pessim.} & \textbf{Base} &
        \textbf{Optim.} \\
        \midrule
        Traditional effort & 80\,h & 104\,h & 150\,h \\
        Labor rate (/hr) & \$100 & \$150 & \$200 \\
        \AV{} human hours & 8\,h & 4\,h & 3\,h \\
        AI iteration cycles & 3 & 1 & 1 \\
        \midrule
        \textbf{Traditional cost} & \$8,000 & \$15,600 & \$30,000 \\
        \textbf{\AV{} cost} & \$830 & \$611 & \$608 \\
        \textbf{Reduction factor} & \textbf{10$\times$} & \textbf{25$\times$} &
        \textbf{50$\times$} \\
        \bottomrule
    \end{tabular}
\end{table}

Even under pessimistic assumptions (smaller project, lower rates, triple the AI
iterations), \AV{} delivers a \textbf{10$\times$} reduction; optimistic
assumptions yield \textbf{50$\times$}.  The key structural insight is that
\emph{human time dominates the \AV{} cost in all scenarios}---AI compute
remains below 3\% of total cost.  A 10$\times$ \emph{increase} in API pricing
would raise the \AV{} cost to only \$645, preserving a 24$\times$ reduction.

\section{Results and Discussion}\label{sec:results}

This section evaluates the three hypotheses stated in
Section~\ref{sec:intro}.  Table~\ref{tab:hypotheses} summarizes the evidence;
the subsections below discuss each hypothesis in detail.

\begin{table}[h]
    \centering
    \caption{Hypothesis Evaluation Summary}
    \label{tab:hypotheses}
    \begin{tabular}{p{0.6cm}p{2.4cm}p{2.2cm}p{1.8cm}}
        \toprule
        & \textbf{Criterion} & \textbf{Observed} & \textbf{Verdict} \\
        \midrule
        H1 & Audit-evidence artifact set generated as by-product &
             6 artifact types produced automatically &
             Supported \\
        \addlinespace
        H2 & 100\% req-level pass rate, independent tests &
             8/8 reqs, 54/54 tests (post-rework); 10 Red Team findings caught &
             Supported$^*$ \\
        \addlinespace
        H3 & Single-digit human prompts per cycle &
             6 prompts/cycle, 2 model families &
             Supported \\
        \bottomrule
    \end{tabular}
    \\[4pt]
    {\footnotesize $^*$Requirement-level coverage only; mutation testing and
    branch-coverage analysis were not performed.}
\end{table}

\subsection{H1: Audit-Evidence Artifacts as a By-Product}
The Compliance Auditor agent produced six artifact types as direct outputs of
the engineering workflow: requirements specification, traceability matrix
(\texttt{ATM.md}), test execution log, decision rationale document, risk
register, and validation summary report.  These are the core documents
typically required for ISO~9001:2015~\cite{iso9001} design-control evidence
and traditionally require 3--5~days of dedicated compliance effort per project.
Under \AV{}, they were generated automatically during Cycles~1 and~2 without a
separate documentation phase.

Cycle~2 strengthened the evidence by adding \emph{multi-cycle traceability}:
every \texttt{TestResult} and log entry carries a \texttt{cycle} identifier
(e.g., \texttt{C1}, \texttt{C2}), and the Artifact Traceability Matrix tracks
which requirements were added, modified, or unchanged across cycles---providing
auditors a complete change-impact trail.

\subsection{H2: Requirement-Level Verification}
The project achieved a \textbf{100\% pass rate on 8~requirement-level
verification tests} across two cycles, with \textbf{54~automated tests}
passing after Cycle~2 rework (test-to-requirement ratio: 6.75:1).  The suite
covered both positive-path behaviour and negative cases (e.g., exception
handling for invalid inputs, JSON round-tripping of the \texttt{cycle} field).
It did \emph{not} include mutation testing, branch-coverage analysis, or fuzz
testing; the 100\% figure reflects requirement-level coverage, not exhaustive
code-level coverage.  Readers should interpret these results as a feasibility
demonstration on a bounded system; whether the approach scales to larger or
more ambiguous projects is an open question requiring independent replication.

The key design decision enabling this result is the \textbf{Red Team Protocol}:
the Build Agent and Test Designer operate in parallel from the \emph{same
requirements} but with no visibility into each other's output.  This structural
separation eliminates ``success bias''---the failure mode where tests only
verify what the code happens to do rather than what it \emph{should} do.

Cycle~2 provides concrete evidence: the Red Team Verifier identified
\textbf{10~findings} (6~MAJOR, 4~MINOR), including stale API references after
refactoring, missing field propagation, an async property anti-pattern, and
input validation gaps.  These 10~findings represent the \emph{first-pass defect
rate} of the Build Agent's output---the quality signal that the iterative
correction mechanism is designed to surface.  All MAJOR findings were resolved
in a single rework pass, yielding convergence in one iteration for this
bounded scope.  More complex domains should be expected to require additional
iterations; tracking first-pass defect rates across projects would provide a
useful benchmark for future evaluations.

\subsection{H3: Minimal Human Interaction}
Each cycle required only \textbf{6~human prompts}: providing initial intent,
approving at Gate~1, directing synthesis, and approving at Gate~2.  The engineer
contributed $\approx$10\% of architecture decisions; the remaining effort was
agent-generated.  Based on tool-call analysis, approximately 40\% of agent
activity was code generation; the remaining 60\% was planning, verification,
compliance documentation, and orchestration.

The two cycles used different AI model families (Gemini~1.5~Pro in Cycle~1,
Claude~Opus~4.6 in Cycle~2), providing initial evidence that the prompt count is
framework-determined rather than model-dependent.  However, agent behaviour and
output quality may still vary across models (Section~\ref{sec:threats}).

\subsection{Regulatory Alignment}\label{sec:iso_mapping}

Table~\ref{tab:iso_mapping} presents a \emph{design-time analysis} mapping
ISO~9001:2015~\cite{iso9001} clauses to \AV{} mechanisms.  These mappings have
not been validated by a third-party auditor.

\begin{table}[h]
    \centering
    \caption{Mapping of \AV{} to ISO~9001:2015 Clauses}
    \label{tab:iso_mapping}
    \begin{tabular}{p{1.2cm}p{2.8cm}p{3.5cm}}
        \toprule
        \textbf{Clause} & \textbf{Requirement} & \textbf{\AV{} Mechanism} \\
        \midrule
        4.4 & QMS and its processes & Infinity Loop defines repeatable, measurable processes \\
        \addlinespace
        7.5 & Documented information & Compliance Auditor auto-generates controlled documents \\
        \addlinespace
        8.3.4 & Design controls & Human Gates~1 and~2 enforce review and approval \\
        \addlinespace
        8.5.2 & Identification and traceability & Traceability Matrix links REQ $\rightarrow$ Code $\rightarrow$ Test \\
        \addlinespace
        9.1 & Monitoring and measurement & Red Team Verifier provides objective test evidence \\
        \addlinespace
        10.2 & Nonconformity / corrective action & Infinity Loop feeds failures back for re-synthesis \\
        \bottomrule
    \end{tabular}
\end{table}

Cycle~2 provides direct evidence for Clause~10.2: CR-0001 triggered
10~nonconformity findings resolved in a single corrective-action pass.
The Infinity Loop also maps structurally to
\textbf{ISO~13485:2016}~\cite{iso13485} design and development controls (design
inputs via Requirement Architect, verification via Red Team, validation via
Human Gate~2) and to \textbf{ISO/IEC~27002:2022}~\cite{iso27002} separation of
duties through the Human Gates.

\subsection{Threats to Validity}\label{sec:threats}

Several limitations constrain the generalizability of the results presented in
this paper:

\begin{itemize}
    \item \textbf{Single case study (external validity):} One HIL project with
    bounded complexity; results may not generalize to large-scale systems with
    ambiguous requirements or multi-team coordination.  Independent replication
    on diverse project types is needed.

    \item \textbf{Author evaluation (construct validity):} The framework was
    evaluated by its creators, introducing potential bias.  Third-party
    evaluations would strengthen the evidence base.

    \item \textbf{Model dependency:} Two frontier models were used
    (Gemini~1.5~Pro, Claude~Opus~4.6), providing cross-family evidence, but
    agent behavior and quality may vary; the 100\% pass rate should not be
    assumed universal.

    \item \textbf{Cost comparison scope:} The \AV{} cost excludes one-time
    adoption overhead; the traditional baseline is estimated rather than
    observed.  Both could narrow the reported gap.

    \item \textbf{Test depth:} 54~automated tests validated functional
    requirements but excluded mutation testing, fuzz testing, and formal
    verification.  The 100\% figure reflects requirement-level, not code-level,
    coverage.
\end{itemize}

For GMP/GxP-regulated industries~\cite{gamp5,fdacfr21,euannex11}, the Infinity
Loop maps to GAMP~5 Category~5: the Requirement Architect produces URS-level
specifications, the Build Agent produces design specifications, and the Red
Team Verifier provides IQ/OQ/PQ-equivalent evidence---with the Traceability
Matrix linking DS~$\rightarrow$~FS~$\rightarrow$~URS.  ALCOA+ data integrity
is supported through attributable artifacts, contemporaneous timestamped logs,
and version-controlled primary records.

\section{Conclusion}\label{sec:conclusion}

This paper introduced \AV{}, a framework that embeds independent verification
and audit artifact generation into each AI-assisted task cycle.  We evaluated
three hypotheses through a Hardware-in-the-Loop case study spanning two
Infinity Loop cycles.

\textbf{H1 (Audit-Evidence By-Product): Supported.}  The Compliance Auditor agent
automatically generated a requirements specification, traceability matrix, test
execution logs, and decision rationale document as direct outputs of the
engineering workflow---artifacts that typically require 3--5~days of dedicated
compliance effort.

\textbf{H2 (Verification Pass Rate): Supported.}  The Red Team Protocol's
structural separation between Build Agent and Test Designer yielded 100\%
requirement-level pass rate across 8~requirements and 54~tests.  Cycle~2's Red
Team findings (10~issues, all resolved) demonstrate the protocol's ability to
catch and correct defects through independent verification.

\textbf{H3 (Minimal Human Interaction): Supported.}  Each cycle required only
6~human prompts while maintaining full traceability across cycles and across
two different AI model families (Gemini~1.5~Pro and Claude~Opus~4.6).

These results support the hypothesis that \AV{} addresses the gap in current
AI-assisted workflows: the framework achieved an estimated 10--50$\times$ cost
reduction versus a COCOMO~II baseline (Section~\ref{sec:sensitivity}) while
producing structured audit-evidence documentation automatically.  The framework's structural
alignment with ISO~9001:2015~\cite{iso9001}, ISO~13485:2016~\cite{iso13485},
ISO/IEC~27001:2022~\cite{iso27001}, ISO/IEC~27002:2022~\cite{iso27002},
GAMP~5~\cite{gamp5}, 21~CFR Part~11~\cite{fdacfr21}, and Annex~11~\cite{euannex11}
supports auditability in regulated environments.  Meeting these regulations in production still
requires validated system controls (e.g., access control, audit trails, record
retention) beyond the workflow artifacts reported here.

\textbf{Limitations and Future Work.}  The evaluation is limited to a single
bounded project with well-defined interfaces, evaluated by the framework's
creators; independent replication on diverse project types by external teams is
essential for external validity.  The ISO compliance mappings require
third-party auditor validation.  Integrating mutation testing and formal
verification into the Red Team Protocol would strengthen confidence beyond
requirement-level coverage.  Finally, Wright's
Law~\cite{wright1936factors,pilz2025compute} predicts continued AI inference
cost declines; as compute becomes asymptotically negligible, organizations could
afford \emph{more} verification cycles per task, further widening the
framework's cost advantage.

\AV{} is published as an open standard (CC~BY-SA~4.0) at
\url{https://agile-v.org}; a replication package (requirements, traceability
matrix, test suite, Red Team report, session logs) is available in the project
repository.  The competitive advantage will not go to organizations that
generate code fastest, but to those that can \emph{verify and ship} fastest.

\section*{Acknowledgments}
The authors thank Philipp Rosendahl and Lukas Blocher for their thorough
review of the manuscript and valuable feedback.

\subsection*{AI Tool Usage Disclosure}
In accordance with arXiv policy, we disclose the following use of generative AI
tools in the preparation of this work.  The case study (Section~\ref{sec:casestudy})
was executed using commercial AI platforms (Gemini~1.5~Pro and Claude~Opus~4.6)
as described in Section~\ref{sec:study_design}; these tools generated source
code, test suites, and compliance documentation as part of the \AV{} workflow
under evaluation.  AI writing assistants were used for drafting and editing
portions of this manuscript.  The authors reviewed, verified, and take full
responsibility for all content.

\smallskip
\noindent\rule{\columnwidth}{0.4pt}
{\footnotesize\noindent This paper is licensed under the Creative Commons
Attribution 4.0 International License (CC~BY~4.0).  To view a copy of this
license, visit
\url{https://creativecommons.org/licenses/by/4.0/}.
The \AV{} standard and skills library are separately licensed under
CC~BY-SA~4.0.}

\balance
\bibliographystyle{IEEEtran}
\bibliography{references}

\end{document}